\documentclass[nofootinbib]{revtex4}
\usepackage{CJK,upgreek,fancyhdr}
\usepackage{graphicx}
\usepackage{latexsym}
\usepackage{amsmath}
\newcommand{\half}{\frac{1}{2}}
\newcommand{\pa}{\partial}

\newcommand{\nn}{\nonumber\\}

\def\be{\begin{equation}}
\def\ee{\end{equation}}
\def\bea{\begin{eqnarray}}
\def\eea{\end{eqnarray}}

\begin{document}
\begin{CJK*}{UTF8}{gbsn}

\hfill USTC-ICTS-17-14

\title{Conformal invariant cosmological perturbations via the covariant approach:  multicomponent universe}

\author{Yunlong Zheng$^{1,2}$}
\author{Yicen Mou$^{3}$}
\author{Haomin Rao$^{3}$}
\author{Mingzhe Li$^{3}$}
\email{limz@ustc.edu.cn}
\affiliation{$^{1}$Department of Physics, Nanjing University, Nanjing 210093, China}
\affiliation{$^{2}$Joint Center for Particle, Nuclear Physics and Cosmology, Nanjing 210093, China}
\affiliation{$^{3}$Interdisciplinary Center for Theoretical Study, University of Science and Technology of China, Hefei, Anhui 230026, China}


\begin{abstract}
In recent years there has been a lot of interest in discussing frame dependences/independences of the cosmological perturbations under  the conformal transformations.
This problem has previously been investigated in terms of the covariant approach for a single component universe, and it was found that the covariant approach is very powerful to pick out the perturbative variables which are both gauge and conformal invariant. In this work, we extend the covariant approach to a universe with multicomponent fluids. We find that similar results can be derived, as expected. In addition, some other interesting perturbations are also identified to be conformal invariant, such as entropy perturbation between two different components.

\end{abstract}

\maketitle

\hskip 1.6cm PACS number(s): 98.80.-k \vskip 0.4cm

\section{Introduction}

Cosmological perturbation theory constitutes the cornerstone of our knowledge to understand the origin and evolution of the large-scale-structure in our universe. This theory had been plagued by the gauge issue: not all the perturbative variables which appear in this theory correspond to real and physical perturbations. The uncertainties originate from arbitrary choices of the correspondence between the real inhomogeneous and anisotropic universe and the background Friedmann-Lema\^itre-Robertson-Walker (FLRW) spacetime. One solution to the gauge problem is fixing a gauge at the beginning. Another is to circumvent it by focusing on the gauge invariant quantities. There are two approaches to find the gauge invariant perturbations. In the conventional coordinate approach, gauge invariant perturbations can be constructed as combinations of gauge-dependent metric and matter perturbations in a specified coordinate, as first done by Bardeen \cite{Bardeen:1980kt} and reviewed in Refs.~\cite{Kodama:1985bj,Mukhanov:1990me,Ma:1995ey}. Another approach is the so-called covariant approach which was developed in Refs.~\cite{Ellis:1989jt,Ellis:1989ju,Bruni:1992dg,Dunsby:1991xk}, based on earlier works by Ehlers \cite{Ehlers}, Hawking \cite{Hawking:1966qi} and Ellis \cite{Ellis:1971pg}. With this approach, all the perturbative variables are covariantly defined and gauge invariant perturbations can be selected out in terms of the Stewart-Walker Lemma \cite{Stewart:1974uz}. For example, for linear perturbation theory, according to this lemma, a covariantly defined variable which vanishes at the background is automatically a gauge invariant perturbation. The advantage of the covariant approach is that all the gauge invariant perturbations have clear geometric and physical meanings. One can refer to Refs. \cite{Challinor:1998xk,Hu:2004xd,Vitenti:2013hda,Osano:2006ew} for more applications and discussions of this approach.

Besides the gauge issue, the problem of whether the cosmological perturbations depend on the frame has attracted much interest in recent years. As we know, when considering those theories in which the gravity is different from general relativity or in which matter couples to gravity non-minimally, such as the Brans-Dicke theory \cite{Brans:1961sx}, $f(R)$ theory \cite{Sotiriou:2008rp,Nojiri:2017ncd}, Galileon theory \cite{Nicolis:2008in,Deffayet:2009wt,Deffayet:2011gz} and so on, we are confronted with the problem of frame choice. Theoretically, there are infinitely many frames which can be used. Different frames are related by the conformal transformations (Weyl rescalings), $\tilde{g}_{ab}=\Omega^2 g_{ab}$. The two most familiar frames used in scalar-tensor theories are the Jordan frame and the Einstein frame. Conventionally, 
in the Jordan frame, matter is minimally coupled to the metric but the action for gravity contains a non-minimal coupling of a scalar field to the Ricci scalar. However, after transforming to the Einstein frame, the action for gravity becomes the Einstein-Hilbert one, but matter is non-minimally coupled to the scalar field. Although variables  change from one frame to another, the physics should be equivalent at least at the classical level. Especially, the observables should be frame independent or conformal invariant. It is known from studies with the coordinate approach that some key cosmological perturbations \cite{Catena:2006bd,Chiba:2013mha,Gong,Chiba,Prokopec,Kubota:2011re} are frame independent or conformal invariant at the linear and non-linear order. One example of a conformal invariant variable is the Weyl tensor. Another famous example is the co-moving curvature perturbation $\zeta$ \cite{Bardeen:1983qw} in single scalar field inflation models. Some implications of the equivalence between different frames in the early universe were discussed in Refs. \cite{Bezrukov:2007ep,Li,Piao,Qiu:2012ia,Domenech:2015qoa,other1,other2,Domenech:2016yxd,Cai:2016gjd,Wetterich:2015ccd,Li:2016gqh,Cai:2015yza,Bahamonde:2017kbs,Bahamonde:2016wmz}.

In our previous work \cite{Li:2015hga}, we studied the problem of frame (in)dependences of cosmological perturbations via the covariant approach, focusing on a cosmology model with a single component. We have investigated how the common perturbative variables (covariantly defined) change under the conformal transformation, and have shown that the covariant approach is very convenient and powerful to pick out the cosmological perturbations which are both gauge and conformal invariant. In this paper, we will generalize the method developed in our previous work and apply it to a universe with multiple components. As we know,  our universe contains many species, including baryons, photons, neutrinos, dark matter, dark energy and so on. Even in studies on the primordial universe, such as inflation, bouncing and emergent universe, models with multiple fields are frequently proposed and investigated. Hence this generalization is necessary. For a multicomponent universe, in many cases, each component can be treated approximately as a perfect fluid. We will use this approximation through out this paper.

This paper is organized as follows. In Section II, we will briefly review the covariant approach and define covariant and gauge invariant variables which characterize the velocity and density perturbations in a multicomponent fluid medium. In Section III the transform rules of various variables under the conformal transformation will be derived and some conformal invariant perturbations will be identified. In Section IV the links of the covariant approach to the coordinate approach will be presented, and we can see what forms those conformal invariant perturbations in the covariant approach take in the coordinate approach. In Section V we apply our results to an example where the gravity is modified. In Section VI we present our conclusions. 

\section{The covariant approach}

At the first step of the covariant approach, one chooses a preferred family of world lines representing the motion of typical observers (fundamental observers) in the universe. The four-velocity $u^a=dx^a/d\lambda$ (tangent to these world lines)  is timelike, future-directed and unit. This is used to define the projection tensor into the tangent three-space orthogonal to $u^a$, 
\be
h_{ab}=g_{ab}+u_au_b~,~{\rm with}~h^a_{~b}h^b_{~c}=h^a_{~c}~,~h_a^{~b}u_b=0~.
\ee
Then the first covariant derivative of the four-velocity is decomposed as follows,
\be
\nabla_b u_{a}=\omega_{ab}+\sigma_{ab}+\frac{1}{3}\Theta h_{ab}-a_au_b~,
\ee
where $\omega_{ab}$ is the antisymmetric vorticity tensor  with $\omega_{ab}u^b=0$, $\sigma_{ab}$ is the symmetric and traceless shear tensor with $\sigma_{ab}u^b=0$ and $\sigma^a_{~a}=0$, $\Theta\equiv \nabla_a u^a$ is the local expansion rate, and $a_{a}\equiv u^b\nabla_b u_{a}$ is the acceleration vector and also orthogonal to the velocity, $a_a u^a=0$. The vorticity and shear magnitudes are defined by $\omega^2\equiv(1/2)\omega_{ab}\omega^{ab}$, $\sigma^2\equiv(1/2)\sigma_{ab}\sigma^{ab}$.
For our purposes, it is useful to introduce a local scale factor $S=e^{\alpha}$, where $\alpha$ is the integration of $\Theta$ along the flow lines with respect to the proper time,
\be\label{alpha}
\alpha\equiv \frac{1}{3}\int d\lambda \Theta~,
\ee
which is defined up to an integration constant.

The matter sector is described by its energy-momentum tensor. For a single component perfect fluid, the fluid velocity is identified with the four-velocity of the fundamental observers $u^a$. Thus we have
\be\label{EMT}
T_{ab}=\rho u_au_b+ph_{ab}~,
\ee
where $\rho=T_{ab}u^au^b$ is the proper density and $p=(1/3)h^{ab}T_{ab}$ is the pressure.
In the case of a scalar field, we usually  define a four-velocity as
\be\label{velocity of scalar}
	u_a=-\frac{\nabla_a\phi}{\sqrt{-\nabla^b\phi\nabla_b\phi}},
\ee
then the energy-momentum tensor of the scalar field has the same form of that of perfect fluid, and the covariant approach can be applied in a similar way.

As we mentioned in the previous section, the real universe in general contains several components. For simplicity we will assume each component is a perfect fluid. 
Then for each component there is a projection tensor $h_{ab}^{(m)}=g_{ab}+u_a^{(m)}u_b^{(m)}$  associated with its four-velocity $u^{a}_{(m)}$, where $m$ represents the $m$th component. Correspondingly we have  the vorticity $\omega_{ab}^{(m)}$, shear $\sigma_{ab}^{(m)}$, acceleration $a_a^{(m)}$, local expansion rate $\Theta_{(m)}$, and $\alpha^{(m)}$ through the decomposition of the first derivative of $u^{a}_{(m)}$, and  density $\rho^{(m)}$ and pressure $p^{(m)}$ read from the energy-momentum tensor $T_{ab}^{(m)}$, which has the following form,
\be\label{each EMT}
	T_{ab}^{(m)}=\rho_{(m)} u_a^{(m)}u_b^{(m)}+p_{(m)}h_{ab}^{(m)}~. 
\ee
Furthermore, one can also define a total velocity $u^{a}$ for all the components so that the total energy-momentum tensor is
\be\label{momentum tensor}
T_{ab}=\rho u_au_b+ph_{ab}+q_{a}u_{b}+u_{a}q_{b}+\pi_{ab}~,
\ee
where $\rho=T_{ab}u^au^b$ and $p=(1/3)h^{ab}T_{ab}$ is total density and pressure, and $q_{a}=-h_a^{~c}T_{cd}u^d$ and  $\pi_{ab}=h_a^{~c}h_b^{~d}T_{cd}-\frac{1}{3}h_{ab}(h^{cd}T_{cd})$ is the total energy flux and anisotropic pressure.

In terms of the Stewart-Walker lemma: the quantities which vanish in the FLRW universe are gauge invariant perturbations. Some basic gauge invariant perturbations can be found easily through the above discussion:
\begin{itemize}
  \item The vorticity, shear and acceleration:
    \be
      a_{a}^{(m)}~,~\omega_{ab}^{(m)}~,~\sigma_{ab}^{(m)}.
    \ee
  \item The matter tensor components:
    \be
       q_a^{(m)}\equiv-h_a^{c}T_{cd}^{(m)}u^d~,~\pi_{ab}^{(m)}\equiv h_a^{~c}h_b^{~d}T_{cd}^{(m)}-\frac{1}{3}              h_{ab}(h^{cd}T_{cd}^{(m)}).
    \ee
  \item The relative velocity:
    \be
      u_a^{(m)}-u_a^{(n)}~\mathrm{or}~ u_a^{(m)}-u_a~.
    \ee
  \item The electric and magnetic parts of the Weyl tensor (contraction with either $u_a$ or $u_a^{(m)}$ will give the same result at the linear order):
  \be
  E_{ab}=C_{acbd}u^c u^d~,~H_{ab}={1\over 2} C_{aecd}u^e \eta^{cd}_{~~~bf}u^f~.
  \ee
\end{itemize}

Other gauge invariant perturbations can be obtained from the spatial gradients of various scalar quantities. For one component fluid we can define gauge invariant quantities such as \cite{Ellis:1989jt,Langlois:2005ii,Langlois:2005qp}
\be
X_a=D_a \rho~,~Y_a=D_a p~,~Z_a=D_a\Theta~,~W_a= D_a\alpha~,~D_a\phi~,
\ee
where the derivative $D_a\equiv h_a^{~b}\nabla_b$ is the projection of the covariant derivative into the tangent three-space.
In the case of the multicomponent fluids, to define quantities that characterize the spatial variation of the density $\rho^{(m)}$, pressure $p^{(m)}$, expansion rate $\theta^{(m)}$, and $\alpha^{(m)}$ of the individual components, we have two choices. We could either define the spatial derivative of each component with respect to the total matter rest frame,
\be
X_a^{(m)}=D_a \rho^{(m)}~,~Y_a^{(m)}=D_a p^{(m)}~,~Z_a^{(m)}=D_a\Theta^{(m)}~,~W_a^{(m)}= D_a\alpha^{(m)}~,~D_a\phi^{I}~,
\ee
where the derivative $D_a\equiv h_a^{~b}\nabla_b$ is the projection of the covariant derivative into the tangent three-space orthogonal to the total matter velocity $u^a$, or we could define gradients for the individual components with respect to the matter rest frame of the components themselves,
\be
^{\star} X_a^{(m)}=D_a^{(m)} \rho^{(m)}~,~^{\star} Y_a^{(m)}=D_a^{(m)} p^{(m)}~,~^\star Z_a^{(m)}=D_a^{(m)}\Theta^{(m)}~,~^\star W_a^{(m)}= D_a^{(m)}\alpha^{(m)}~,~D_a^{(I)}\phi^{I}~,
\ee
where $D_a^{(m)}\equiv h_a^{(m)b}\nabla_b$ is the spatial gradient orthogonal to the $m$th fluid velocity $u^a_{(m)}$.
Note that the quantities like $X_a^{(m)},Y_a^{(m)},W_a^{(m)}$ contain information about both the $m$th component fluid and the total fluid. Furthermore, $D_a^{(I)}\phi^{I}$ exactly vanishes according to the definition of the four velocity of the scalar field.

\section{Conformal invariant perturbations}

Following our previous work, we will first derive the transform rules of the covariantly defined  variables under conformal transformation and then pick up the perturbations which are both gauge and conformal invariant. In this paper we will attach more importance to the multicomponent matter sector than the curvature variables, because the latter has been discussed in our previous work \cite{Li:2015hga}. 

Under the conformal transformation $\tilde{g}_{ab}=\Omega^{2}g_{ab}$, we have
\bea\label{con1}
& &d\tilde{\lambda}=\Omega d\lambda~,\tilde{u}^a_{(m)}=\Omega^{-1} u^a_{(m)}~,
~\tilde{u}_a^{(m)}=\Omega u_a^{(m)}~,~\tilde{h}_a^{(m)b}=h_a^{(m)b}~.
\eea
With these relations, we can immediately find the transform rules of the kinematical variables
\bea\label{con_congruence}
& &\tilde{\omega}_{ab}^{(m)}=\Omega\omega_{ab}^{(m)}~,~\tilde{\sigma}_{ab}^{(m)}=\Omega\sigma_{ab}^{(m)}~,
~{\rm or}~\tilde{\omega}^{(m)}=\Omega^{-1}\omega^{(m)}~,~\tilde{\sigma}^{(m)}=\Omega^{-1}\sigma^{(m)}~,\nonumber\\
& &\tilde{\Theta}^{(m)}=\Omega^{-1}\Theta^{(m)}+3\Omega^{-2}\dot\Omega~,~\tilde{a}_a^{(m)}=a_a^{(m)}+D_a^{(m)}\ln\Omega~,\nonumber\\
& &\tilde{\alpha}^{(m)}=\alpha^{(m)}+\ln \Omega~,~\tilde{S}^{(m)}=\Omega S^{(m)}~,~~
\tilde{W}_a^{(m)}=W_a^{(m)}+ D_a\ln \Omega~,^{\star}\tilde{W}_a^{(m)}=^{\star}W_a^{(m)}+ D_a^{(m)}\ln \Omega~.
\eea
In addition, from the definition of the energy-momentum tensor through the variation of the action with respect to the metric, one can obtain the following transform
\be
\tilde{T}_{ab}^{(m)}=\Omega^{-2}T_{ab}^{(m)}~.
\ee
Thus the energy density and pressure of matter have the conformal weight $4$, i.e.,
$\tilde{\rho}^{(m)}=\Omega^{-4}\rho^{(m)}~,~\tilde{p}^{(m)}=\Omega^{-4} p^{(m)}$,
and the following ratios change as
\begin{align}\label{con_density}
&\frac{\tilde{X}_a^{(m)}}{\tilde{\rho}^{(m)}}=\frac{X_a^{(m)}}{\rho^{(m)}}-4D_a\ln\Omega~,~\frac{\tilde{Y}_a^{(m)}}{\tilde{p}^{(m)}}
=\frac{Y_a^{(m)}}{p^{(m)}}-4D_a\ln\Omega~.\nn
&\frac{^{\star}\tilde{X}_a^{(m)}}{\tilde{\rho}^{(m)}}=\frac{^{\star}X_a^{(m)}}{\rho^{(m)}}-4D_a^{(m)}\ln\Omega~,~\frac{^{\star}\tilde{Y}_a^{(m)}}{\tilde{p}^{(m)}}
=\frac{^{\star}Y_a^{(m)}}{p^{(m)}}-4D_a^{(m)}\ln\Omega~.
\end{align}
Furthermore, the energy flux contributed by the $m$th fluid, defined as $q_a^{(m)}\equiv-h_a^{c}T_{cd}^{(m)}u^d$, is of conformal weight 3. So, we immediately obtain the following gauge and conformal invariant quantity:
 \be
   \frac{q_a^{(m)}/S}{\rho^{(m)}+p^{(m)}}
 \ee

With the above transform rules, we obtain the following covariant quantities which are both gauge and conformal invariant.
\begin{itemize}
  \item Electric and magnetic parts of the Weyl tensor \cite{Li:2015hga}:
    \be\label{inv1}
    E_{ab}~,~H_{ab}
    \ee
  \item The perturbations of the $m$th component:
    \be
    \frac{\omega_{ab}^{(m)}}{S},~\frac{\sigma_{ab}^{(m)}}{S},~^{\star} W_a^{(m)}-a_a^{(m)}~,~\frac{^{\star}X_a^{(m)}}{\rho^{(m)}}+4^{\star}W_a^{(m)}~,
    ~\frac{^{\star}Y_a^{(m)}}{p^{(m)}}+4^{\star}W_a^{(m)}~,~\frac{^{\star}X_a^{(m)}}{\rho^{(m)}}-\frac{^{\star}Y_a^{(m)}}{p^{(m)}}~.\label{inv2}
    \ee
   \item The quantities related to $m$th component and total velocity:
    \be
    \frac{X_a^{(m)}}{\rho^{(m)}}+4W_a^{(m)}~,~\frac{Y_a^{(m)}}{p^{(m)}}+4W_a^{(m)}~,~\frac{X_a^{(m)}}{\rho^{(m)}}-\frac{Y_a^{(m)}}{p^{(m)}}~,\frac{q_a^{(m)}/S}{\rho^{(m)}+p^{(m)}}~.\label{inv3}
    \ee
  \item Relative quantities between the $m$th and $n$th fluids:
     \be\label{inv4}
     ~\frac{u_a^{(m)}-u_a^{(n)}}{S},~W_a^{(m)}-W_a^{(n)},~\frac{X_a^{(m)}}{\rho^{(m)}}-\frac{X_a^{(n)}}{\rho^{(n)}},
     ~\frac{Y_a^{(m)}}{p^{(m)}}-\frac{Y_a^{(n)}}{p^{(n)}}~.
    \ee
\end{itemize}
As usual, we assume all fluid components share the same four velocity at the FLRW background. The total scale factor $S$ and the scale factor $S_{(m)}$ associated with any single component  will give the same result as far as linear perturbation theory is concerned. We should emphasize here that the results in Eqs. (\ref{inv1}-\ref{inv2}) correspond to those listed in Eq. (22) of our previous work \cite{Li:2015hga} in the case of a single fluid. Namely, they are not completely new. However, the quantities shown in Eqs. (\ref{inv3}-\ref{inv4}) are new gauge and conformal invariant perturbations which are absent in the single fluid model.

We can rewrite some of the conformal invariant entropy perturbations  above  in a more simple and elegant form, for example, 
\be
\frac{X_a^{(m)}}{\rho^{(m)}}-\frac{X_a^{(n)}}{\rho^{(n)}}=\frac{D_a(\rho^{(m)}/\rho^{(n)})}{\rho^{(m)}/\rho^{(n)}}=D_a\ln\left(\frac{\rho^{(m)}}{\rho^{(n)}}\right)
\ee
which obviously represents the spatial derivative of density ratio of two components with respect to the total rest frame. With this form, the physical meaning of such entropy perturbations is clearer. Other quantities, such as $\frac{^{\star}X_a^{(m)}}{\rho^{(m)}}-\frac{^{\star}Y_a^{(m)}}{p^{(m)}},\frac{X_a^{(m)}}{\rho^{(m)}}-\frac{Y_a^{(m)}}{p^{(m)}}$ can be similarly rewritten in this way.

Sometimes, the conformal factor $\Omega$ is not arbitrary, and merely depends on a scalar field $\phi$, as in cases of conventional frame transformations in scalar-tensor theories. Furthermore, the derivative $\nabla_a\phi$ should be timelike if $\phi$ is not vanishing in the FLRW background. In this case, we can slice the universe by the spacelike hypersurfaces $\phi=\rm{Consts}$ and define the total hydrodynamical four velocity $u^a$ normal to these hypersurfaces everywhere. With such defined velocity, it is easy to prove that $D_a\phi=0$ and $D_a\ln\Omega(\phi)=0$. So, according to the transform rules in Eq. (\ref{con_congruence}), one can find that the perturbations $W_a^{(m)},X_a^{(m)}/\rho^{(m)},Y_a^{(m)}/p^{(m)}$ are invariant under the conformal transformations with $\Omega(\phi)$, i.e., $\tilde{W}_a^{(m)}=W_a^{(m)}$, $\tilde{X}_a^{(m)}/\tilde{\rho}^{(m)}=X_a^{(m)}/\rho^{(m)}$,~$\tilde{Y}_a^{(m)}/\tilde{p}^{(m)}
=Y_a^{(m)}/p^{(m)}$, though they are not invariant for a general conformal factor. 

In the next section we will focus on the conformal invariant perturbations, especially those listed in Eqs.~(\ref{inv3}-\ref{inv4}), and see what  forms they have in the coordinate approach.

\section{Links to the coordinate approach}

\subsection{General conformal transformation and invariant variables}

As shown in our previous work \cite{Li:2015hga}, in the coordinate approach all the gauge invariant vector and tensor perturbations are conformal invariant. This is because the conformal factor is a scalar field and its inhomogeneity cannot affect the vector and tensor perturbations, at least up to the linear order. Hence, we will only consider the scalar perturbations in the rest of this paper. With the coordinate approach, when perturbed metric is considered, the line element takes the form: 
\be\label{metric}
  ds^2=a^2\{-(1+2A)d\eta^2+2B_{,i} d\eta dx^i+[(1-2\psi)\gamma_{ij}+2E_{|ij}]dx^idx^j\},
\ee
where $A,~B, ~\psi,~E$ denote the perturbations and the subscript $|ij$ represents second order covariant derivative associated with the induced background metric $\gamma_{ij}$, which will be used to lower and raise the indices hereafter. 

According to the normalization of  the velocity $u_a^{(m)}$, the $m$th four-velocity  up to linear order is
\be\label{ua}
 u^a_{(m)}=\left(\frac{1-A}{a},~~\frac{v^{i}_{(m)}}{a}\right)~,
\ee
and
\be
 u_a^{(m)}=\left(-a(1+A), ~~a(B_{,i}+v_i^{(m)})\right)~.
\ee
For scalar perturbation, $v_i^{(m)}$ is generated by a velocity potential $v^{(m)}$, so that $v_i^{(m)}=v_{,i}^{(m)}$.

Applying the above equations we may calculate the covariant quantities listed in the previous sections. All of the quantities will  be calculated up to the linear order. As the conformal invariant quantities listed in Eqs.~(\ref{inv1}-\ref{inv2}) correspond to the quantities listed in Eq.~(22) of \cite{Li:2015hga}, except that there are multicomponent fluids here, they are not totally new. So we directly skip the calculations of the following gauge and conformal invariant perturbations:
\be
E_{ab}~,~H_{ab}~,~\frac{\omega_{ab}^{(m)}}{S}~,~\frac{\sigma_{ab}^{(m)}}{S}~,~^{\star} W_a^{(m)}-a_a^{(m)}~,~\frac{^{\star}X_a^{(m)}}{\rho^{(m)}}+4^{\star}W_a^{(m)}~,
    ~\frac{^{\star}Y_a^{(m)}}{p^{(m)}}+4^{\star}W_a^{(m)}~,~\frac{^{\star}X_a^{(m)}}{\rho^{(m)}}-\frac{^{\star}Y_a^{(m)}}{p^{(m)}}~.
\ee
 
Now we calculate the quantity $W_a^{(m)}$. Its spatial component is
 \be\label{wi}
   W_i^{(m)}= \partial_i[\mathcal{H}(B+v)-\psi+\frac{1}{3}\int d\eta \Delta (v^{(m)}+E')]~,
 \ee
where $v$ and $v^{(m)}$ represent  velocity perturbations of total matter and the $m$th component respectively. From this result we can see it is indeed different from Eq. (36) of our previous work \cite{Li:2015hga}, as expected.

Now we expand the gauge invariant  perturbation $X_a^{(m)}/\rho^{(m)}~\mathrm{and}~Y_a^{(m)}/p^{(m)}$ to the linear order, giving the non-vanishing spatial components
 \bea\label{density perturbation}
   \frac{X_i^{(m)}}{\rho^{(m)}}&=&\left[\frac{\delta\rho^{(m)}}{\rho^{(m)}}+\frac{\rho'^{(m)}}{\rho^{(m)}}(B+v)\right]_{,i}~,\nonumber\\
   \frac{Y_i^{(m)}}{p^{(m)}}&=&\left[\frac{\delta p^{(m)}}{p^{(m)}}+\frac{p'^{(m)}}{p^{(m)}}(B+v)\right]_{,i}~.
 \eea
We recognize that $\mathcal{R}^{(m)}_1=\frac{\delta\rho^{(m)}}{\rho^{(m)}}+\frac{\rho'^{(m)}}{\rho^{(m)}}(B+v)$ and $\mathcal{R}_2^{(m)}=\frac{\delta p^{(m)}}{p^{(m)}}+\frac{p'^{(m)}}{p^{(m)}}(B+v)$ are  the comoving density and pressure contrast of the $m$th fluid with respect to the total matter rest frame. They are gauge invariant but generally not conformal invariant. Then we expand the gauge and conformal invariant quantities $X_a^{(m)}/\rho^{(m)}+4 W_a^{(m)}$, $Y_a^{(m)}/p^{(m)}+4 W_a^{(m)}$ and $X_a^{(m)}/\rho^{(m)}-Y_a^{(m)}/p^{(m)}$ to linear order and get the non-vanishing components
 \bea
   \frac{X_i^{(m)}}{\rho^{(m)}}+4W_i^{(m)}&=&\left[\frac{\delta\rho^{(m)}}{\rho^{(m)}}-4\psi+(\frac{\rho'^{(m)}}{\rho^{(m)}}+4\mathrm{H})(B+v)+\frac{4}{3}\int d\eta \Delta (v^{(m)}+E')\right]_{,i}~, \\
   \frac{Y_i^{(m)}}{p^{(m)}}+4W_i^{(m)}&=&\left[\frac{\delta p^{(m)}}{p^{(m)}}-4\psi+(\frac{p'^{(m)}}{p^{(m)}}+4\mathrm{H})(B+v)+\frac{4}{3}\int d\eta \Delta (v^{(m)}+E')\right]_{,i}~,\\
   \frac{X_i^{(m)}}{\rho^{(m)}}-\frac{Y_i^{(m)}}{p^{(m)}}&=&\left[\frac{\delta\rho^{(m)}}{\rho^{(m)}}-\frac{\delta p^{(m)}}{p^{(m)}}+(\frac{\rho'^{(m)}}{\rho^{(m)}}-\frac{p'^{(m)}}{p^{(m)}})(B+v)\right]_{,i}~.
 \eea
which tell us that $\mathcal{R}^{(m)}=\frac{\delta\rho^{(m)}}{\rho^{(m)}}-4\psi+(\frac{\rho'^{(m)}}{\rho^{(m)}}+4\mathcal{H})(B+v)$, $\mathcal{R}^{(m)}_3=\frac{\delta p^{(m)}}{p^{(m)}}-4\psi+(\frac{p'^{(m)}}{p^{(m)}}+4\mathrm{H})(B+v)$ and $\mathcal{R}^{(m)}_4=\frac{\delta\rho^{m}}{\rho^{m}}-\frac{\delta p^{m}}{p^{m}}+(\frac{\rho'^{(m)}}{\rho^{(m)}}-\frac{p'^{(m)}}{p^{(m)}})(B+v)$ are both gauge and conformal invariant. The variable $\mathcal{R}^{(m)}$ is more meaningful when the universe is dominated by radiation fluid, as discussed in Ref.~\cite{Li:2015hga}. In that case, it becomes
 \be\label{uniform density curvature perturbation}
  \frac{\mathcal{R}^{(m)}}{4}=-\psi+\frac{\delta\rho^{(m)}}{3(\rho^{(m)}+p^{(m)})}~,
 \ee
which is the curvature perturbation on uniform-density hypersurfaces, used extensively in cosmological perturbation theory, and the quantity $\mathcal{R}^{(m)}$ is now only related to the $m$th component. Note when the fluid has a constant equation of state (EOS), the variable $\mathcal{R}^{(m)}_4$ will be zero . Actually one can prove that in the case of constant EOS, the term $\frac{X_a^{(m)}}{\rho^{(m)}}-\frac{Y_a^{(m)}}{p^{(m)}}$ will exactly vanish.

From the previous section we know that $q_a^{(m)}/[S(\rho^{(m)}+p^{(m)})]$ is both gauge and conformal invariant. Its non-vanishing component is
 \be
   \frac{q_i^{(m)}/S}{\rho^{(m)}+p^{(m)}}=[v^{(m)}-v]_{,i}
 \ee
This means $v^{(m)}-v$ is both gauge and conformal invariant, as is $v^{(mn)}\equiv v^{(m)}-v^{(n)}$. 

Now we calculate the relative quantities between different components.
First we have the relative four-velocity, which is gauge and conformal invariant:
\be
   \frac{u_a^{(m)}-u_a^{(n)}}{S}=(0,~~v_{,i}^{(m)}-v_{,i}^{(n)})~.
\ee
Again, we get that $v^{(m)}-v^{(n)}$ is gauge and conformal invariant.
The non-vanishing component of gauge and conformal invariant quantity $W_a^{(m)}-W_a^{(n)}$ is
 \be
   W_i^{(m)}-W_i^{(n)}=\left[\frac{1}{3}\int d\eta \Delta(v^{(m)}-v^{(n)})\right]_{,i}~,
 \ee
and once again we get the invariant quantity $v^{(mn)}$.

Next we calculate the relative  density  perturbation between two components. Its spatial component is
\be
   X_i^{(m)}-X_i^{(n)}=\left[(\frac{\delta\rho^{(m)}}{\rho^{(m)}}-\frac{\delta\rho^{(n)}}{\rho^{(n)}})+
   (\frac{\rho'^{(m)}}{\rho^{(m)}}-\frac{\rho'^{(n)}}{\rho^{(n)}})(B+v)\right]_{,i}=\left[\mathcal{R}^{(m)}_5\right]_{,i}~.
\ee
This means $\mathcal{R}^{(m)}_5$ is gauge and conformal invariant. One can calculate  $Y_a^{(m)}-Y_a^{(n)}$ in a similar way.

Now focus on the case of multiple scalar fields. The scalar field is usually invariant when the theory transforms  from one frame to another frame  in  scalar-tensor theories such as Brans-Dicke theory \cite{Brans:1961sx}, Galileon theory \cite{Nicolis:2008in,Deffayet:2009wt,Deffayet:2011gz}, and so on.
For a scalar field $\phi^{I}$ with zero conformal weight, which means the scalar itself is conformal invariant, we have a gauge and conformal invariant quantity $D_a\phi^{I}$. Its non-vanishing spatial component up to linear order is
\be
  ~D_i\phi^{I}=[\delta\phi^{I}+\phi^{I'}(B+v)]_{,i}=[\delta\phi^{I(gi)}+\phi^{I'}(v+E')]_{,i}~,
\ee
which means $\delta\phi^{I(gi)}=\delta\phi^{I}+(B-E')\phi^{I'}$ is conformal invariant. This is consistent with our previous result in Ref. \cite{Li:2015hga}. One can immediately find another composite conformal invariant quantity 
\be
 \frac{D_i\phi^{I}}{\phi^{I'}}-\frac{D_i\phi^{J}}{\phi^{J'}}=\left[\frac{\delta\phi^{I}}{\phi^{I'}}-\frac{\delta\phi^{J}}{\phi^{J'}}\right]_{,i}
\ee
where $S^{IJ}=\frac{\delta\phi^{I}}{\phi^{I'}}-\frac{\delta\phi^{J}}{\phi^{J'}}$   represents the frequently used entropy perturbation between two scalar fields   
\cite{Gordon:2000hv,Li:2013hga,Qiu:2014apa}. This means $S^{IJ}$ must be conformal invariant, which is consistent with the calculation of the conformal invariant quantity $v^{IJ}$. 

\subsection{Restricted conformal transformations and invariant quantities}

We know via the covariant approach that (to linear order) the vector and tensor perturbations, and the sums of the scalar perturbations $v^{(m)}+E'$ ,$v^{(m)}-v^{(n)}$ and $\Psi+\Phi$ and so on in the coordinate approach are both gauge and conformal invariant, whatever the conformal factor is. Now we will consider the restricted conformal transformation in which the conformal factor only depends on a (timelike) scalar field $\phi$, i.e., $\Omega=\Omega(\phi)$. This often happens in scalar-tensor theories. In these models the Jordan frame and Einstein frame are related by such conformal transformations.  In this case, it is convenient to define the total four-velocity as
 \be
   u_{a}=-\frac{\nabla_{a}\phi}{\sqrt{-\nabla_b\phi\nabla^b\phi}}~.
 \ee
It is then obvious that $u_a$ is orthogonal to the hypersurfaces $\Omega=\rm{const}$, and, as we stressed at the end of the previous section, the perturbations $W_a^{(m)}$, $X_a^{(m)}/\rho^{(m)}$, $Y_a^{(m)}/p^{(m)}$ and $D_a\phi^{I}/\phi^{I}$ themselves are both gauge and conformal invariant. Firstly, the expansion of $W_a$ gives the following  conformal invariant quantity
 \be\label{comoving curvature}
    \zeta=-\psi+\mathcal{H}(B+v)~.
 \ee
The variable $\zeta$ is the curvature perturbation on uniform-$\phi$ hypersurfaces, and in the case of  single fluid it is just the co-moving curvature perturbation which takes an important role in cosmological perturbation theory. The expansion of $X_a^{(m)}/\rho^{(m)}$ and $Y_a^{(m)}/p^{(m)}$ tell us that the co-moving density perturbation $\mathcal{R}_1^{(m)}$ and pressure perturbation $\mathcal{R}_2^{(m)}$ are conformal invariant. The expansion of $D_a\phi^{I}/\phi^{I}$ gives another conformal invariant quantity
 \be
    \mathcal{R}_6^{I}=\frac{\delta\phi^{I}}{\phi^{I}}+\frac{\phi^{I'}}{\phi^{I}}(B+v).
 \ee

 From the above discussions we know that $\zeta$, $\mathcal{R}_1^{(m)}$, $\mathcal{R}_2^{(m)}$ and  $\mathcal{R}_6^{I}$  are conformal invariant.
In term of the identifications, one find
 \be
   u_i=-a(\frac{\delta\phi}{\phi'})_{,i}=a(B+v)_{,i}~,~B+v=-\frac{\delta\phi}{\phi'}~.
 \ee
We  get the explicit form of the curvature perturbation on uniform-$\phi$ hypersurfaces as:
 \be
   \zeta=-\psi+\mathcal{H}(B+v)=-\psi-\mathcal{H}\frac{\delta\phi}{\phi'}~,
 \ee
 and other invariants:
 \bea
   \mathcal{R}_1^{(m)}&=&\frac{\rho'^{(m)}}{\rho^{(m)}}(\frac{\delta\rho^{(m)}}{\rho'^{(m)}}-\frac{\delta\phi}{\phi'})~,\nonumber\\
   \mathcal{R}_2^{(m)}&=&\frac{p'^{(m)}}{p^{(m)}}(\frac{\delta p^{(m)}}{p'^{(m)}}-\frac{\delta\phi}{\phi'})~.
 \eea
The invariant $\mathcal{R}_1^{(m)}$ ($\mathcal{R}_2^{(m)}$) is proportional to the entropy perturbation between the density (pressure) of the $m$th component and the field.
For a scalar field other than $\phi$ which is related to the conformal factor, we have the invariant $\mathcal{R}_6^{I}$. Its detailed form is
 \be
   \mathcal{R}_6^{I}=\frac{\phi'^{I}}{\phi^{I}}(\frac{\delta\phi^{I}}{\phi'^{I}}-\frac{\delta\phi}{\phi'}),
 \ee
which is proportional to the entropy perturbation between two scalar fields $\phi^{I}$ and $\phi$.

\section{One example}
Let us take the model of $f(R)$ gravity with a scalar field as an example. This model was considered in Ref. \cite{Qiu:2014apa}, and there the action is
\be\label{fR}
	S=\int d^4x\sqrt{-\tilde{g}}\left[\frac{f(\tilde{R})}{2}+\mathcal{L}_s \right]
\ee
where $f$ is an arbitrary function and $\mathcal{L}_s=-\tilde{g}^{\mu\nu}\pa_\mu\chi\pa_\nu\chi-\mathcal{\nu}(\chi)$ is the Lagrangian for the matter field $\chi$. 
For convenience, we use variables with tildes to refer to those in the Jordan frame, while their counterparts without tildes are those in the Einstein frame. 
As we know, the theory  \eqref{fR} can be rewritten into the Brans-Dicke form by the Legendre transformation
\be\label{BD}
	S_{J}=\int d^4x\sqrt{-\tilde{g}}\left[\frac{\varphi\tilde{R}}{2}-U(\varphi)+\mathcal{L}_s \right]
\ee
where $\varphi\equiv F,~~U(\varphi)=F\tilde{R}-f(\tilde{R})$.
 Here $F$ is defined as $F\equiv\pa f/\pa \tilde{R}$. There are two scalars in this action, and one of them is non-minimally coupled to gravity. By a conformal transformation we can shift to the Einstein frame in which the gravity is minimally coupled. The metric in the Einstein frame connects to the original metric as
\be\label{conformal2}
	g_{\mu\nu}=\Omega^2\tilde{g}_{\mu\nu}
\ee
where the conformal factor is $\Omega=\sqrt{\varphi}$. The action in the Einstein frame is
\be\label{EF} 
	S_E=\int d^4x\sqrt{-g}\left[\frac{R}{2}-\frac{3}{4}\frac{\nabla_\mu\varphi\nabla^\mu\varphi}{\varphi^2}- \frac{1}{2\varphi}\nabla_\mu\chi\nabla^\mu\chi-V(\varphi,\chi)\right],
\ee
where
\be 
	V(\varphi,\chi)=\varphi^{-2}(U(\varphi)+\nu(\chi)).
\ee
Furthermore, one can define the new variable $\phi=-\sqrt{6}/2\ln\varphi$ to simplify the non-minimal kinetic term. After field redefinitions, we have
\be 
	S_E=\int d^4x\sqrt{-g}\left[\frac{R}{2}-\half\nabla_\mu\phi\nabla^\mu\phi-\half e^{\frac{2}{\sqrt{6}}\phi}\nabla_\mu\chi\nabla^\mu\chi-V(\phi,\chi)\right].
\ee

Under the conformal transformation \eqref{conformal2}, the action \eqref{BD} in the Jordan frame becomes Eq.~\eqref{EF}. Two scalar fields $\varphi$ and $\chi$ are invariant, and thus have zero conformal weight. In addition, the conformal factor is the function of the scalar field $\varphi$, which is exactly the case of restricted conformal transformation discussed above. According to our analysis, the entropy perturbation between two scalar fields, 
\be 
	\delta s=\frac{\delta\chi}{\chi'}-\frac{\delta\phi}{\phi'}
\ee
is both gauge and conformal invariant. Another invariant quantity is the famous curvature perturbation on uniform-$\varphi$ hypersurfaces (sometimes also called the co-moving curvature perturbation), 
\be
	\zeta=-\psi-\mathcal{H}\frac{\delta\varphi}{\varphi'}=-\psi-\mathcal{H}\frac{\delta\phi}{\phi'}.
\ee

Note that the co-moving curvature perturbation defined by
\be
	\mathcal{R}=-\psi-\frac{\mathcal{H}}{\rho+P}\delta q,
\ee
is not conformal invariant in this case,  although it is indeed invariant in the case of scalar-tensor theory with only one scalar degree of freedom.  Here $\delta q$ is defined to satisfy the relation $\pa_i\delta q=\delta T^0_i$. Our result is consistent with the calculation in Ref. \cite{Qiu:2014apa}, where the difference of co-moving curvature perturbations in two frames is derived.
However, this dose not mean that these two frames are not equivalent. The frame dependent comoving curvature perturbation $\mathcal{R}$ only means it represents different variables in different frames, and the equations of motions are also different. 

\section{Conclusions}

Conformal transformations connect one frame to another, and are frequently used in scalar-tensor theories, including equivalent modified gravities.  As the physical observables should be not only gauge invariant but also independent of frame, it is important to find those quantities which are both gauge and conformal invariant. In our previous paper, we revisited the problem of conformal invariances of cosmological perturbations using the covariant approach, in which the geometric and physical meanings of the perturbative variables are very clear. In this work we extend the covariant formalism to a universe with multicomponent fluids. Besides some similar results to our previous work, we find some other interesting perturbations which are also conformal and gauge invariant, such as the covariantly defined quantities listed in Eqs.(\ref{inv3}-\ref{inv4}). These quantities represent the entropy perturbations between different physical variables or different components. When translating by the language of the coordinate approach, we find quantities which are invariant under general conformal transformation, such as
 $\mathcal{R}^{(m)}, \mathcal{R}^{(m)}_3, \mathcal{R}^{(m)}_4, \mathcal{R}^{(m)}_5, v^{(m)}-v^{(n)}, \delta\phi^{(I)gi}$, and $\frac{\delta\phi^{(I)}}{\phi'^{(I)}}-\frac{\delta\phi^{(J)}}{\phi'^{(J)}}$ for the zero weight field. We also showed the second kind conformal invariant quantities, which are invariant under the restricted conformal transformation, such as $\mathcal{R}^{(m)}_1,\mathcal{R}^{(m)}_2,\mathcal{R}_6^{I}$, which represent the entropy perturbations between the density (pressure, or another scalar) and the scalar field the conformal factor depends on. 

\section{Acknowledgement}

This work is supported in part by NSFC under Grant No. 11422543 and No. 11653002.

{}

\clearpage
\end{CJK*}
\end{document}